# Analytic expression of the temperature increment in a spin transfer torque nanopillar structure

Chun-Yeol You,[1] Seung-Seok Ha,[1] and Hyun-Woo Lee[2]

[1]Department of Physics, Inha University, Incheon 402-751, Korea
[2]PCTP and Department of Physics, Pohang University of Science and Technology, Pohang 790-784, Korea

Ed. - Highlight - Please verify the technical meaning of the edited document.

The temperature increment due to the Joule heating in a nanopillar spin transfer torque system is investigated. We obtain a time dependent analytic solution of the heat conduction equation in nanopillar geometry by using the Green's function method after some simplifications of the problem. While Holm's equation is applicable only to steady states in metallic systems, our solution describes the time dependence and is also applicable to a nanopillar-shaped magnetic tunneling junction with an insulator barrier layer. The validity of the analytic solution is confirmed by numerical finite element method simulations and by the comparison with Holm's equation.

PAC: 85.75.-d, 72.25.Ba, 72.25.Pn, 72.15.Eb

I. INTRODUCTION

The spin transport torque (STT) in magnetic multilayers is an exciting research subject and is responsible for interesting phenomena such as current induced magnetization switching, the emission of microwaves, and current induced domain wall motion.[1~5] The spin dynamics of multilayers have been intensively studied under strong non-equilibrium conditions subject to a spin polarized current with a high current density, where the Joule heating that accompanies the current becomes an important factor. The temperature of the magnetic system is important for the quantitative analysis of the spin dynamics,[6,7,8] and is also related to the stability of the STT device. Although the Joule-heating-induced temperature increment has been estimated in a few studies,[9,10] these estimations are rather case-specific and a more systematic investigation is therefore needed. In this respect, Holm's equation[11] is very valuable, since it relates the maximum temperature increment of arbitrary-shaped contacts with their electrical resistance. The validity of Holm's equation has been tested by various experiments over the past few decades, however it has a few shortcomings. First, it is applicable only when the Wiedemann-Franz law holds in all parts of the system. Thus, when the system contains an insulator barrier in the current path, which is the case in magnetic tunnel junctions (MTJs), the equation is not applicable. Furthermore, this equation provides only the steady state temperature, while information on the temporal variation of the temperature is desired for certain device applications. In this paper, we solve the time-dependent heat conduction equation for a nanopillar geometry, which may contain an insulating layer in the current path. After some reasonable approximations, a time dependent analytic solution for the heat conduction equation is obtained by employing the Green's function method.[12,13] Though the calculation scheme bears some resemblance to that used for nanowire studies,[14,15] it turns out that the temperature increments in nanopillars and nanowires have noticeable differences. We also test and confirm the validity of the approximations by performing numerical calculations using commercial finite element method software.[16] The result is also compared with that given by Holm's equation. This paper is organized as follows. In Sec. II, we derive the analytic expression for the temperature increment for a homogeneous nanopillar, and a comparison with Holm's equation is given in Sec. III. In Secs. IV and V, we apply the analytic expression to spin-valve and MTJ nanopillars, respectively, and compare the results with those obtained from the numerical simulations. In Sec. VI, we compare the temperature increments for the nanowire and nanopillar systems. A more detailed comparison between the calculation results and reported experimental data is given in Sec. VII. The conclusions are presented in Sec. VIII.

## II. ANALYTIC EXPRESSION FOR THE TEMPERATURE INCREMENT IN A NANOPILLAR

In order to solve the heat conduction equation for the nanopillar geometry, we make the following approximations. First, the heat generated in the electrodes is ignored and the heat is assumed to be generated only inside of the nanopillar. This approximation is reasonable, since the nanopillar is generally much smaller than the electrodes and the current density in the nanopillar is much higher than that in the electrodes. Furthermore, the nanopillar includes high resistivity materials such as a barrier layer and/or antiferromagnetic layer. Thus, most of the power consumption occurs within the nanopillar. For the sake of simplicity, semi-infinite electrodes are considered below. Second, we ignore the heat conduction through the insulator layer surrounding the nanopillar (not to be confused with the insulator layer within the current path of an MTJ), as shown in Fig. 1 (a). To verify this approximation, we perform numerical simulations of the heat conduction equation, taking explicit account of the heat conduction through the surrounding insulator. Commercial finite element software is used for the simulations. $SiO_2$, $Si_3N_4$, and a fictitious insulator material whose thermal conductivities are 20, 1.4 and 0.01 W/m K, respectively, are used as the surrounding insulator layers. The temperature increments, $\Delta T$, at the center of the nanopillar generated by a 10 ns pulse are plotted in Fig. 2 as a function of time. Note that the results of the two real insulators are very similar to that for the fictitious insulator material with negligible heat conductivity, thereby confirming the validity of the second assumption. Thus, the insulator may be replaced by a vacuum with zero thermal conductivity. Third, the nanopillar is assumed to be mirror symmetric with respect to the $z=0$ plane. Although real nanopillars do not have mirror symmetry, explicit numerical calculations (Secs. IV and V) suggest that the breaking of the symmetry does not significantly affect the result.

Once mirror symmetry is imposed, the calculation can be considerably simplified. The geometry can be simplified as shown in Fig. 1 (b) and the temperature increment at the center of the nanopillar can be calculated in two steps. In the first step, we calculate the temperature increment, $\Delta T$, at the interface between the nanopillar and the electrodes. In the second step, the temperature difference between the interface and the center of the nanopillar is calculated. The calculation of the first step is similar to the nanowire problem.[14] Following the procedure described in Ref. [14], the original nanopillar problem is simplified to the structure shown in Fig. 1 (c), where the half nanopillar is regarded as a heat source generating the downward heat flow, $g_J(x,y,t)$, into the bottom electrode. If we assume uniform Joule heating inside of the nanopillar, the true lateral profile of the generated heat is given by

$$g_J^{True}(x,y,t) = \frac{P_0}{2}\theta\left(1 - \frac{x^2}{r_x^2} - \frac{y^2}{r_y^2}\right), \tag{1}$$

where $\theta(x)$ is a step function. However, the mathematical treatment of this realistic heat source is rather complex, so that we take the approximation of a Gaussian heat source for the sake of mathematical simplicity, as we did in the nanowire case.[14,15,17]

$$g_J(x,y,t) = \frac{P_0/2}{\pi r_x r_y}\exp\left(-\frac{x^2}{r_x^2} - \frac{y^2}{r_y^2}\right), \tag{2}$$

where $r_x = \alpha r_{x0}$, $r_y = \alpha r_{y0}$, and $r_{x0}$ and $r_{y0}$ are the radii of the nanopillar in the *x* and *y* directions, respectively. The lateral profiles of the realistic and Gaussian heat sources are illustrated in Fig. 1 (c). During this approximation, we introduce an adjustable parameter, $\alpha \sim 1$, which is of the order of unity. The physical meaning of $\alpha$ becomes clear when we compare Eq. (2) and (1) and, from Fig. 1 (c), it can be seen that $\alpha$ is an adjustable parameter related to the effective radius of the heat source. Let $P_0/2$ denote the Joule heating generated inside of the half nanopillar due to the current density, *J*, where $P_0$ is given by

$$P_0 = RI^2 = \frac{d(\pi r_{x0} r_{y0} J)^2}{\sigma_{NP} \pi r_{x0} r_{y0}} = \frac{\pi r_{x0} r_{y0} d J^2}{\sigma_{NP}}. \tag{3}$$

Here, $\sigma_{NP}$ and $d$ are the electrical conductivity and height of the nanopillar, respectively. Let us focus on the most interesting place, the center of the interface between the nanopillar and the bottom electrode. To make the connection with the nanowire calculation[14] straightforward, we choose this center point as the origin of the system in the first step of the calculation. For simplicity, we assume axial symmetry, viz.

$r = \alpha r_0$, $r_0 = r_{x0} = r_{y0}$. Using a similar procedure to that described in Ref. [14], the temperature at the center of the interface, $T_E(t) \equiv T(x=0, y=0, z=0, t)$, is readily obtained for a step pulse of pulse duration time, $t_p$:

$$T_E(t) = T_{E0} G(t). \tag{4}$$

where

$$T_{E0} = \frac{P_0/2}{\pi^2 r_0 \alpha K_E} = \frac{r_0 dJ^2}{2\pi\alpha\sigma_{NP}K_E}, \tag{5}$$

$$G(t) = \left( \tan^{-1}\left[\frac{2\sqrt{\pi\mu_E t}}{\alpha r_0}\right] - \theta(t-t_p)\tan^{-1}\left[\frac{2\sqrt{\pi\mu_E (t-t_p)}}{\alpha r_0}\right] \right). \tag{6}$$

Here, $\theta(t)$ is the step function and $K_E$ is the thermal conductivity of the electrode. For $t \gg \tau \equiv \frac{\alpha^2 r_0^2}{4\pi\mu_E}$, $T_E(t \to \infty) = \frac{\pi T_{E0}}{2}$ since $G(t \to \infty) = \pi/2$. Here, the time constant, $\tau$, is about $1.8 \times 10^{-12}$ s for a Cu electrode with $r_0 = $ 50 nm. The physical meaning of $\tau$ will be discussed later.

Next, we calculate $T_{NP}(t)$, the temperature increment at the center of the nanopillar. The heat conduction within the nanopillar is almost one-dimensional, since the heat flux from the nanopillar to the surrounding insulator is negligible, as demonstrated above. For the moment, we ignore the temporal variation and focus only on the steady state profile. At the steady state, the time dependence in the heat conduction equation can be ignored. We build up a one dimensional heat conduction equation,

$$\frac{\partial^2 T_{NP}(z)}{\partial z^2} + \frac{1}{K_{NP}} g(z) = 0. \tag{7}$$

where the heat source $g(z) = \frac{P_0}{2\alpha^2 \pi r_0^2 d}$ and $K_{NP}$ is the thermal conductivity of the nanopillar. Here, we shift the origin of the coordinate system to the center of the nanopillar. The solution of Eq. (7) is easily obtained with the proper boundary conditions: $T_{NP}(z = \pm d/2) = \frac{\pi}{2} T_{E0}$. By assuming mirror symmetry, we can find the solution at $z = 0$.

$$T_{NP}(z=0) = \frac{\pi T_{E0}}{2}\left(1 + \frac{K_E}{4K_{NP}} \frac{d}{\alpha r_0}\right). \tag{8}$$

Equation (8) has no time dependence, because we neglect it during the derivation procedure. According to the numerical study, however, it is found that the time dependences of $T_{NP}(t)$ and $T_E(t)$ are similar to each other, as shown in Fig. 3. Therefore, it is reasonable to approximate the time dependence of $T_{NP}(t)$ as follows,

$$T_{NP}(z=0,t) = T_{E0}\left(1 + \frac{K_E}{4K_{NP}} \frac{d}{\alpha r_0}\right) G(t). \tag{9}$$

This is the central result of our study. We remark that even though Eq. (9) is derived for a homogenous metallic nanopillar, it is applicable to a much wider class of nanopillar systems. For a multilayer stack, for instance, Eq. (9) remains valid after trivial replacements: $\sigma_{NP} \to \sigma_{eff}$, $K_{NP} \to K_{eff}$, where the effective electrical and thermal conductivities $\sigma_{eff}$ and $K_{eff}$, respectively, are defined by

$$d/\sigma_{eff} = d_1/\sigma_1 + d_2/\sigma_2 + \cdots + d_N/\sigma_N, \tag{10}$$

$$d/K_{eff} = d_1/K_1 + d_2/K_2 + \cdots + d_N/K_N. \tag{11}$$

Here, $d_i$, $\sigma_i$, and $K_i$ are the thickness, electrical conductivity, and thermal conductivity for the $i$-th layer, respectively, in the multilayer stack. In later sections, we will also demonstrate the applicability of Eq. (8) to spin valve structures (Sec. IV) and MTJ nanopillars (Sec. V).

A comparison between Eq. (9) and the numerical calculation is depicted in Fig. 4 (a) for the simple nanopillar configuration, where the nanopillar consists of a single metallic material and, thus, the mirror symmetry is respected. In this numerical calculation, both the nanopillar and the top/bottom electrodes are made of Cu and the insulator layer is assumed to be made of $SiO_2$. Both the radius and thickness of the nanopillar are 50 nm and the current density $J = 10^{12}$ A/m$^2$. With the adjustable parameter, $\alpha = 0.885$, the agreement between the numerical result and Eq. (9) is almost perfect. Here, we discuss briefly the physical meaning of $\tau = \frac{\alpha^2 r_0^2}{4\pi\mu_E}$. It is clearly a characteristic time scale in this problem. It must be pointed out that $\tau$ is mainly determined by the diffusion constant of the electrodes, not the nanopillar materials. Since thermal stabilization is achieved when the input heat is balanced by the escaping heat, it is natural that $\tau$ depends on the thermal diffusion constant of the electrodes. Also, the relevant length scale of the heat is the radius of the nanopillar. According to Eq. (9), the temperature will be saturated within a few $\tau$. For example, about 94 % of the temperature increment is achieved within $t = 10\tau$ in the case of semi-infinite electrodes for a full metallic system. A typical metal electrode gives $\tau \sim$ 1~10 ps, so that the temperature of the nanopillar will be saturated in less than a nanosecond. It must be remarked that the typical pulse width for an MTJ is longer than 1 ns. Therefore, the temperature of the nanopillar will always be saturated in a real experimental situation.

## III. HOLM'S EQUATION

In this section, we demonstrate that in the steady state limit ($t \to \infty$), our results are in good agreement with Holm's equation for metallic nanopillars. R. Holm showed that the temperature increment can be expressed as follows:[11]

$$T_{max}^2 = T_0^2 + \frac{3}{4}\left(\frac{eRI}{\pi k_B}\right)^2, \tag{12}$$

where $R$, $I$, $k_B$, $T_0$, and $T_{max}$ are, respectively, the resistance of the system, current, Boltzmann constant, ambient temperature, and the maximum temperature in the nano contact. For a small temperature increment, Eq. (12) can be rewritten as

$$T_{max} \approx T_0 + \frac{3}{8T_{max}}\left(\frac{eRI}{\pi k_B}\right)^2. \tag{13}$$

For a nanopillar structure that consists of a nanopillar and two electrodes,

$$R \approx R_{NP} + R_E = \frac{1}{\sigma_{NP}}\frac{d}{\pi r^2} + \frac{2\beta}{\sigma_E}\frac{r}{\pi r^2} \tag{14}$$

where the second term represents the contribution of the electrodes. Since we assume that the size of the electrodes is semi-infinite, their contribution to the resistance is only limited around the contacts. A rough estimation of this contribution is the area, $\pi r^2$, and length $r$ with a conductivity of $\sigma_E$. Therefore, the resistance of the electrodes is $\frac{2\beta}{\sigma_E}\frac{r}{\pi r^2}$ where we introduce an adjustable parameter, $\beta$, which is of the order of unity. For $R_{NP} \gg R_E$, which is usually the case, the temperature increment can be written as

$$\Delta T = T_{max} - T_0 \approx \frac{3}{2}\left(\frac{e}{\pi k_B}\right)^2\left(\frac{\beta r_0 dJ^2}{\sigma_{NP}\sigma_E T_{max}}\right)\left(1 + \frac{\sigma_E}{4\beta\sigma_{NP}}\frac{d}{r_0}\right). \tag{15}$$

Finally, by using the Wiedemann-Franz law,[18] the electrical conductivity (of the electrode?) can be replaced by its thermal conductivity,

$$\Delta T \approx \frac{\beta r_0 dJ^2}{2\sigma_{NP}K_E}\left(1 + \frac{K_E}{4\beta K_{NP}}\frac{d}{r_0}\right). \tag{16}$$

Equation (16) is identical to the prediction of Eqs. (8) and (9) in the limit of $t \to \infty$ (except for the adjustable parameters $\beta$). This demonstrates that for metallic nanopillars, the long-term behavior of our results agrees with that of Holm's equation.

IV. SPIN-VALVE STRUCTURE

The real nanopillars used in STT applications usually have a spin-valve structure and do not respect mirror symmetry. Here, we test the applicability of Eq. (9) to a real nanopillar structure by numerically calculating the temperature increment in a realistic nanopillar structure and comparing it with that given by Eq. (9). To take a specific case, we consider a spin-valve nanopillar consisting of multilayer stacks such as Ta(10 nm)/IrMn(10 nm)/Co(15 nm)/Cu(10 nm)/Co(2.5 nm)/Ta(2.5 nm).[5] The values of the electrical and thermal conductivities of the various layers are listed in Table I. Figure 4 (b) shows the temperature increment in the spin-valve at the center of the nanopillar, which is inside of the thick Co layer. The solid line represents the result of Eq. (9) with $\alpha = 1.10$. Note that even though the mirror-symmetry assumption does not hold in the spin-valve structure, Eq. (9) still remains reasonably accurate.

V. TUNNELING JUNCTION STRUCTURE

The MTJ structure is the basic building block of STT-MRAM (Magnetoresistive Random Access Memory). Since the MTJ has an insulator barrier layer within the path of the current flow, Holm's equation cannot be used. The heating in the MTJ comes mainly from the heat dissipation of the hot electrons tunneling through the insulator barrier and, for the typical voltage difference applied to the MTJ,[19] the relation between the tunneling current and voltage is not Ohmic. However, the concept of Joule heating is still valid. Here, we test the applicability of Eqs. (8) and (9) to the MTJ structure.

A numerical study for $\Delta T$ in an MTJ has previously been reported.[19] The nanopillar for the MTJ consists of two ferromagnetic layers and a ~1 nm thick insulator barrier layer, and the heat is mainly produced around the insulator barrier layer, due to the hot electrons, rather than in the whole nanopillar structure. We assume semi-infinite top and bottom electrodes, as in the previous case, and perform numerical simulations. We consider an applied voltage of 0.5-V with a current of 200-μA, which are the typical switching conditions for STT MRAM.[20] The corresponding current density is $2.5 \times 10^{10}$ A/m$^2$. The energy dissipation in the system is $P_0 = 10^{-4}$ W. For $T_{E0}$, we can use the first definition in Eq. (5) with the given $P_0$ instead of the second definition in Eq. (4). Therefore, we do not need the electrical conductivity information of the nanopillar. For the effective thermal conductivity concept is used. Here, we assume that the top and bottom ferromagnetic layers are made of Co instead of multilayer stacks, for the sake of simplicity. And we assume that the heat is generated only in the vicinity of the MgO barrier layer due to the hot electrons. It must be mentioned that the numerical results are not sensitive to the size of the dissipation region. We consider a nanopillar with a radius and thickness of 50-nm, and the numerical results of the temperature increment as a function of time with pulse duration $t_p = 10$ ns are plotted in Fig. 5. We also plot Eq. (9) with $\alpha = 0.897$. Surprisingly, the agreement is as good as that obtained in the metallic system, which implies that Eq. (9) is well able to describe the MTJ nanopillar case, where Holm's equation is not valid.

VI. NANOPILLAR VS. NANOWIRE: TEMPERATURE INCREMENT

Here, it must be mentioned that the calculated $\Delta T$ of the nanopillars in the case of the simple metallic, spin-valve or MTJ systems, is only of the order of 1 K in our study. This result is in clear contrast with the result of the nanowire system where the Joule-heating is very severe and the temperature of the nanowire sometimes exceeds the Curie temperature of the system.[21] The typical $\Delta T$ of the nanowire is of the order of several hundred K. In order to explain the large discrepancy between the nanowire and nanopillar cases, we compare the leading terms of Eq. (9) and the nanowire case, $\Delta T_{WIRE} \sim \frac{whJ^2}{\pi \sigma_{NW} K_S}$.[14,15] Here, $w$, $h$, and $\sigma_{NW}$ are the width, thickness, and electrical conductivity of the nanowire, respectively, and $K_S$ is the thermal conductivity of the semi-infinite substrate. The expressions for the nanopillar and nanowire are alike and the geometrical dimensions of the nanopillar and the nanowire are both a few tens of nanometers. However, there are two big differences between the two. The first one is the thermal conductivity. In the nanowire case, what is relevant is the thermal conductivity of the insulating or semi-insulating substrate, while in the nanopillar case, the thermal conductivity of the metallic electrodes is the relevant parameter. Typically, the thermal conductivities of metals are much larger than those of insulating or semi-insulating materials. The second point is the total power consumption. For the nanowire, the total volume of the heat generating part is large, because of the semi-infinite length of the nanowire, while the total volume of the nanopillar is finite. Therefore, even when the input current density is the same for both cases, the total power consumption in the nanopillar is much smaller than that in the nanowire. Furthermore, the nanopillar is in contact with large electrodes, which are good thermal conductors. Therefore, the generated heat can escape easily through the electrodes and it is natural for the $\Delta T$ of the nanopillar to be much smaller than that for the nanowire.

## VII. COMPARISON WITH EXPERIMENTAL DATA

In order to compare our analytic and numerical results with the reported experimental data, we collected the published experimental data from many groups, as shown in Table II.[2,5,22,23,24] The $\Delta T$ values are evaluated from the resistance increments by using the relation $\Delta R/R = \alpha_T \Delta T$ for the metals. Here, $\alpha_T$ is the temperature coefficient of resistivity and $\alpha_T \sim 5\times 10^{-3}$ (K$^{-1}$) for various transition metals.[25] It must be mentioned that the estimation of $\Delta T$ by this method may contain errors, but we believe that the order of magnitude remains correct for metallic systems, since their magnetoresistance is rather small (~1 %).

The typical increment of $\Delta R/R$ due to the temperature increment in metallic systems is about 5 % for $J = 10^{12}$ A/m$^2$, which corresponds to $\Delta T \sim 10$ K. This is much larger than our estimated value for a simple Cu nanopillar ($\Delta T \sim 0.05$), but is in reasonable agreement with the spin-valve result ($\Delta T \sim 4$ K) shown in Figs. 4 (a) and (b). Here, we would like to point out that the electrical and thermal conductivities of Ta, IrMn, and MgO are much smaller than those of typical ferromagnetic and nonmagnetic metals, as tabulated in Table I. According to Eq. (9), the smaller effective electrical and thermal conductivities of the spin-valve are ascribed to the larger $\Delta T$ for the same $J$. Furthermore, it has been pointed out that the experimentally measured resistances of nanopillars are somewhat larger than their bulk values.[26] We introduce a parameter $k$ (=$R_{exp.}/R_{nom.}$), which is the ratio between the experimentally measured and nominal resistances. Because of the unavoidable interface resistances in the multilayer stacks, $k$ = 1~5 is a reasonable range of values.[26] If the electrical conductivity of the nanopillar decreases, the thermal conductivity also decreases, and the temperature is raised further.

While we can extract $\Delta T$ for spin-valve systems from the published experimental data, it is impossible in the case of the MTJ, because the tunneling magnetoresistance is too large (~100 %) in comparison to the changes due to $\Delta T$. In the MTJ case, even though the MgO barrier is not a good thermal conductor, the thickness of MgO is only 1 nm, while that of the antiferromagnetic layer is >10 nm for the spin-valve structure. However, for the MTJ with a large electrical resistance, the power consumption is larger, so that $\Delta T$ will be large. Unfortunately, it is rather difficult to extract $\Delta T$ from the published data by using the relation $\Delta R/R = \alpha_T \Delta T$, since the tunneling magnetoresistance is quite large (~100%) in MTJs and can induce a non-zero $\Delta R$ even when $\Delta T = 0$.

Furthermore, it should be mentioned that we assume semi-infinite top/bottom electrodes in our study, which is not realistic. Especially, this assumption is totally incorrect in real STT-MRAM device structures, where the electrodes are a kind of nanowires and have small volumes. Therefore, our study is not applicable to realistic devices. For device level analyses, some excellent numerical works have previously been published.[27]

## VIII. CONCLUSIONS

In conclusion, we derived a time dependent analytic solution of the temperature increment in a Joule heated nanopillar for the study of the STT in magnetic multilayers. The validity of the analytic expression is confirmed by the comparison between the numerical calculations and Holm's equation. It is also confirmed that the analytic expression is applicable to an MTJ nanopillar that contains an insulating layer in the current path. From the analytic and numerical calculations, we found that the temperature increment of the nanopillar is not large for metallic and low resistance MTJ systems. The validity of the present expression is also confirmed by the comparison with the estimated temperature increment obtained from the published experimental data.

ACKNOWLEDGMENTS

This work was supported by KOSEF (Basic Research Program No. R01-2007-000-20281-0) and the Nano R&D program through the Korea Science and Engineering Foundation funded by the Ministry of Science & Technology (2008-02553). The authors wish to thank Prof. J. Bass for his helpful comments.

1. J. C. Slonczewski, J. Magn. Magn. Mater. **159**, L1 (1996): L. Berger, Phys. Rev. B **54**, 9353 (1996).

2. E. B. Myers, D. C. Ralph, J. A. Katine, R. N. Louie, and R. A. Buhrman, Science **285**, 867 (1999): S. I. Kiselev, J. C. Sankey, I. N. Krivorotov, N. C. Emley, R. J. Schoelkopf, R. A. Buhrman, and D. C. Ralph, Nature **425**, 380 (2003).

3. Y. Tserkovnyak, A. Brataas, G. E. W. Bauer, and B. I. Halperin, Rev. of Mod. Phys. **77**, 1375 (2005).

4. A. Brataas, G. E. W. Bauer, P. J. Kelly, Phys. Rep. **427**, 157 (2006).

5. J. C. Lee, M. G. Chun, W. H. Park, C.-Y. You, S.-B. Choe, W. Y. Yung and K. Y. Kim, J. Appl. Phys. **99**, 08G517 (2006).

6. S. Urazhdin, N. O. Birge, W. P. Pratt, Jr., and J. Bass, Phys. Rev. Lett. **91**, 146803 (2003): A. Fábián, C. Terrier, S. S. Guisan, X. Hoffer, M. Dubey, L. Gravier, and J.-Ph. Ansermet, Phys. Rev. Lett. **91**, 257209 (2003).

7. S. Petit, C. Baraduc, C. Thirion, U. Ebels, Y. Liu, M. Li, P. Wang, and B. Dieny, Phys. Rev. Lett. **98**, 077203 (2007).

8. S. Petit, N. de Mestier, C. Baraduc, C. Thirion, Y. Liu, M. Li, P. Wang, and B. Dieny, Phys. Rev. B **78**, 184420 (2008).

9. G. D. Fuchs, I. N. Krivorotov, P. M. Braganca, N. C. Emley, A. G. F. Garcia, D. C. Ralph, and R. A. Buhrman, Appl. Phys. Lett. **86**, 152509 (2005).

10. G. D. Fuchs, J. A. Katine, S. I. Kiselev, D. Mauri, K. S. Wooley, D. C. Ralph, and R. A. Buhrman, Phys. Rev. Lett. **96**, 186603 (2006).

11. R. Holm, "Electric Contacts; Theory and applications," Springer-Verlag, p.63 (1967).

12. H. S. Carslaw and J. C. Jaeger, "Conduction of Heat in Solids", 2$^{nd}$ Ed. (Clarendon, Oxford, 1959): J. V. Beck, K. D. Cole, A. Haji-Sheikh, and B. Litkouhi, "Heat Conduction Using Green's Functions" (Hemisphere, Washington, DC, 1992).

13. M. K. Loze and C. D. Wright, Appl. Opts. **36**, 494 (1997).

14. C.-Y. You, I. M. Sung, and B.-K. Joe, Appl. Phys. Lett. **89**, 222513 (2006).

15. C.-Y. You and S.-S. Ha, Appl. Phys. Lett. **91**, 022507 (2007).

16. COMSOL Multiphysics, http://www.comsol.com.

[17] S.-S. Ha and C.-Y. You, J. of Magnetics **12**, 7 (2007).

[18] C. Kittel, "Introduction to Solid State Physics," 8th Ed. John Wiley & Sons, Inc. p. 156 (2005).

[19] S.-S. Ha and C.-Y. You, Phys. Stat. Sol. (a) **204**, 3966 (2007).

[20] M. Hosomi, *et al.* IEDM Technical Digest, IEEE International 2005, 459 (2005).

[21] F. Junginger, M. Kläui, D. Backes, U. Rüdiger, T. Kasama, R. E. Dunin-Borkowski, L. J. Heyderman, C. A. F. Vaz, and J. A. C. Bland, Appl. Phys. Lett. **90**, 132506 (2007).

[22] T. Ochiai, *et al.* Appl. Phys. Lett. **86**, 242506 (2005).

[23] M. AlHajDarwish, H. Kurt, S. Urazhdin, A. Fert, R. Loloee, W. P. Pratt, Jr., and J. Bass, Phys. Rev. Lett. **93**, 157203 (2004).

[24] J. Grollier, V. Cros, H. Jaffrès, A. Hamzic, J. M. George, G. Faini, J. B. Youssef, H. Le Gall, and A. Fert, Phys. Rev. B **67**, 174402 (2003).

[25] D. Halliday, R. Resnick, and J. Walker, "Fundamentals of Physics", John Wiely & Sons. Inc. 6th Ed. Table 27-1, p. 619 (2001).

[26] A. Fukushima, H. Kubota, A. Yamamoto, Y. Suzuki, and S. Yuasa, IEEE Trans. **MAG 41**, 2571 (2005). : A. Fukushima, K. Yagami, A. Tulapurkar, Y. Suzuki, H. Kubota, A. Yamamoto, and S. Yuasa, Jap. J. Appl. Phys. **44**, L12 (2005).

[27] D. H. Lee and S. H. Lim, Appl. Phys. Lett. **92**, 233502 (2008).

Figure Captions

Fig. 1 (a) Schematic diagram of the nanopillar structure for the heat conduction equation. The outward heat flux from the nanopillar is assumed to flow only to the top and bottom electrodes. The radius and thickness of the nanopillar are $r$ and $d$, respectively. With the assumption of mirror symmetry, the problem can be simplified, as shown in (b). For the purpose of calculating the temperature at the interface between the nanopillar and the bottom electrode, the heat source can be treated as if it were confined at the interface, as shown in (c). (d) The temperature increment $T_{NP}(0,t)$ at the center of the nanopillar can be obtained from the one-dimensional heat conduction equation with $T_E(t)$ [Eq. (3)] as the boundary condition at the two ends of the nanopillar.

Fig. 2. The temperature increment $T(t)$ as a function of time at the center of the nanopillar with $r = 50$ nm and $d = 50$ nm for a pulse duration time of 10 ns and current density $J = 10^{12}$ A/m$^2$. For both the nanopillar and the electrodes, the material parameters for Cu are used, while for the insulator, three different sets of parameters are used; SiO$_2$ ($K$=1.4 W/mK), Si$_3$N$_4$ ($K$=20 W/mK), and a fictitious insulator material with negligible heat conductivity ($K$=0.01 W/mK). The tiny difference between the results for the three insulators indicates that the heat conduction through the insulator layer is negligible.

Fig. 3 Numerical calculations of temperature increment for $T_E(t)$ and $T_{NP}(t)$ as a function of time. A Cu nanopillar ($r = d = 50$ nm) with semi-infinite top/bottom Cu

electrodes is considered for $J = 10^{12}$ A/m$^2$.

Fig. 4 Comparisons between numerical calculations and analytic solutions. The blue rectangles and the solid line represent the numerical results and Eq. (9), respectively. (a) For Cu nanopillar with $\alpha = 0.885$. (b) For a realistic spin valve nanopillar with $\alpha = 1.10$. The effective electrical and thermal conductivities are used in the analytic calculations.

Fig. 5 Comparisons between numerical calculations and analytic solutions for the MTJ nanopillar case. The nanopillar consists of Co(24.5 nm) /MgO (1 nm) /Co(24.5 nm) with semi-infinite top and bottom electrodes. The blue rectangles and the red solid lines represent the numerical and analytic results with $\alpha = 0.897$, respectively.

Table I. The electrical and thermal conductivities for the various materials.

| Materials | $\sigma\,(\Omega m)^{-1}$ | $K$ (W/(m K)) |
|:---:|:---:|:---:|
| Ta | $6.5\times 10^5$ | 58 |
| Co | $1.6\times 10^7$ | 692 |
| Cu | $5.9\times 10^7$ | 400 |
| IrMn | $6.8\times 10^5$ | 35.6 |
| MgO |  | 45 |

Table II. The value of $\Delta T$ calculated from the resistance changes for various published data. The value of $k$ is estimated from $R_{nom} = \rho d / A$ with the bulk resistivity. In some cases, the thickness of the nano-pillar is not clear, so that we estimate $k$ (=$R_{exp}/R_{nom}$) roughly.

| Pillar structure | R (Ω) | Pillar size (nm×nm) | J ($10^{12}$ A/m$^2$) | k | $\Delta R/R$ | $\Delta T$ (K) | Ref. |
|---|---|---|---|---|---|---|---|
| Cu/Co/Cu/Co/Cu/Pt | 18.95 | 130×70 | 1.12 | 18.2 | 0.04 | 9.8 | [2] |
| Ta/IrMn/CoFe/Cu/CoFe | 0.53 | 100×150 | 3.3 | 1.39 | 0.04 | 9.7 | [5] |
| Ta/Cu/CoFe/Cu/CoFe/Cu/Ta | 3.87 | 110×270 | 0.98 | 15.6 | 0.018 | 4.2 | [22] |
| Fe/Cr/Fe | 3.94 | 70×130 | 0.98 | 7.48 | 0.025 | 5.9 | [23] |
| Ni/Cu/Py | 2.27 | 70×130 | 1.4 | 3.4 | 0.038 | 9.3 | [23] |
| Co/Cu/Co | 0.244 | 200×600 | 5.31 | 18.2 | 0.02 | 4.7 | [24] |

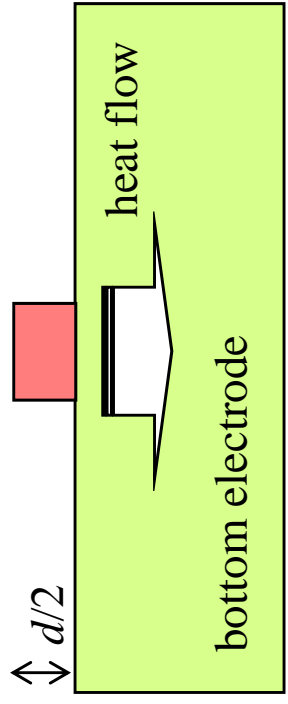
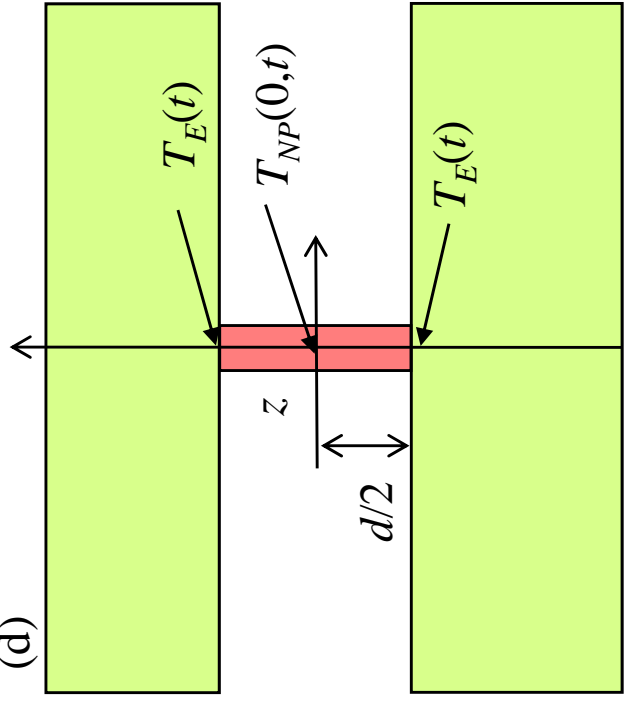
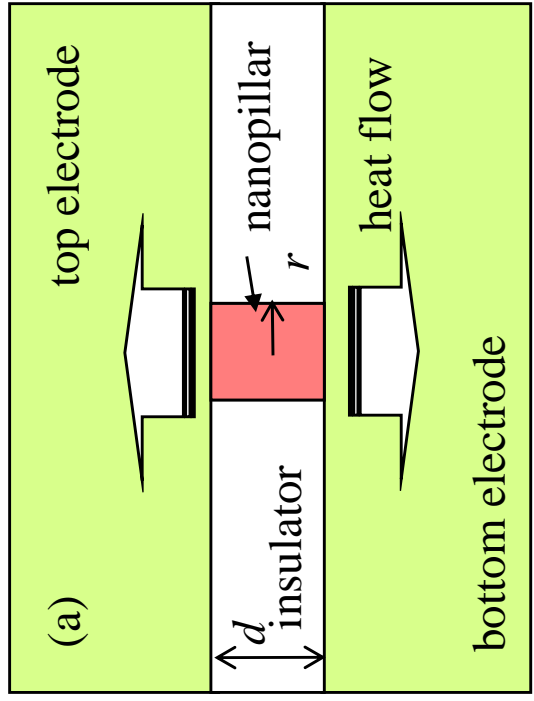
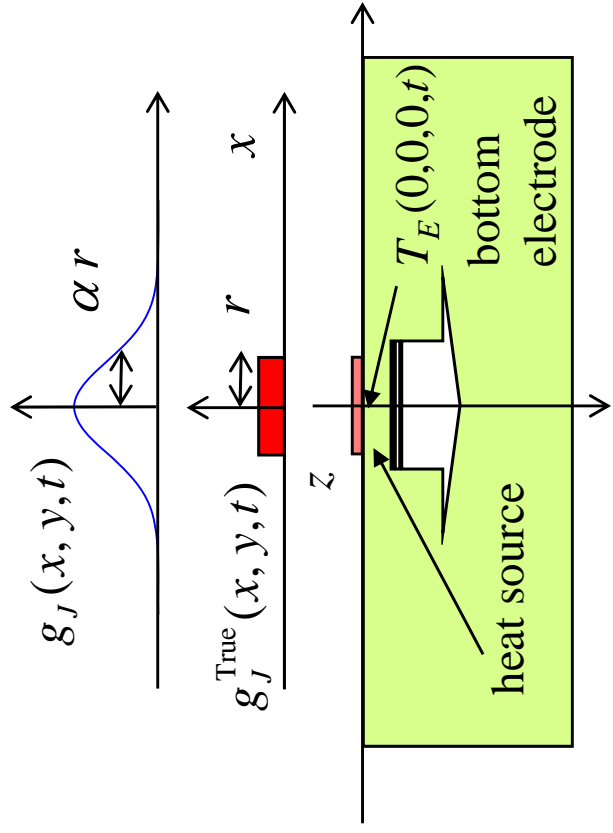

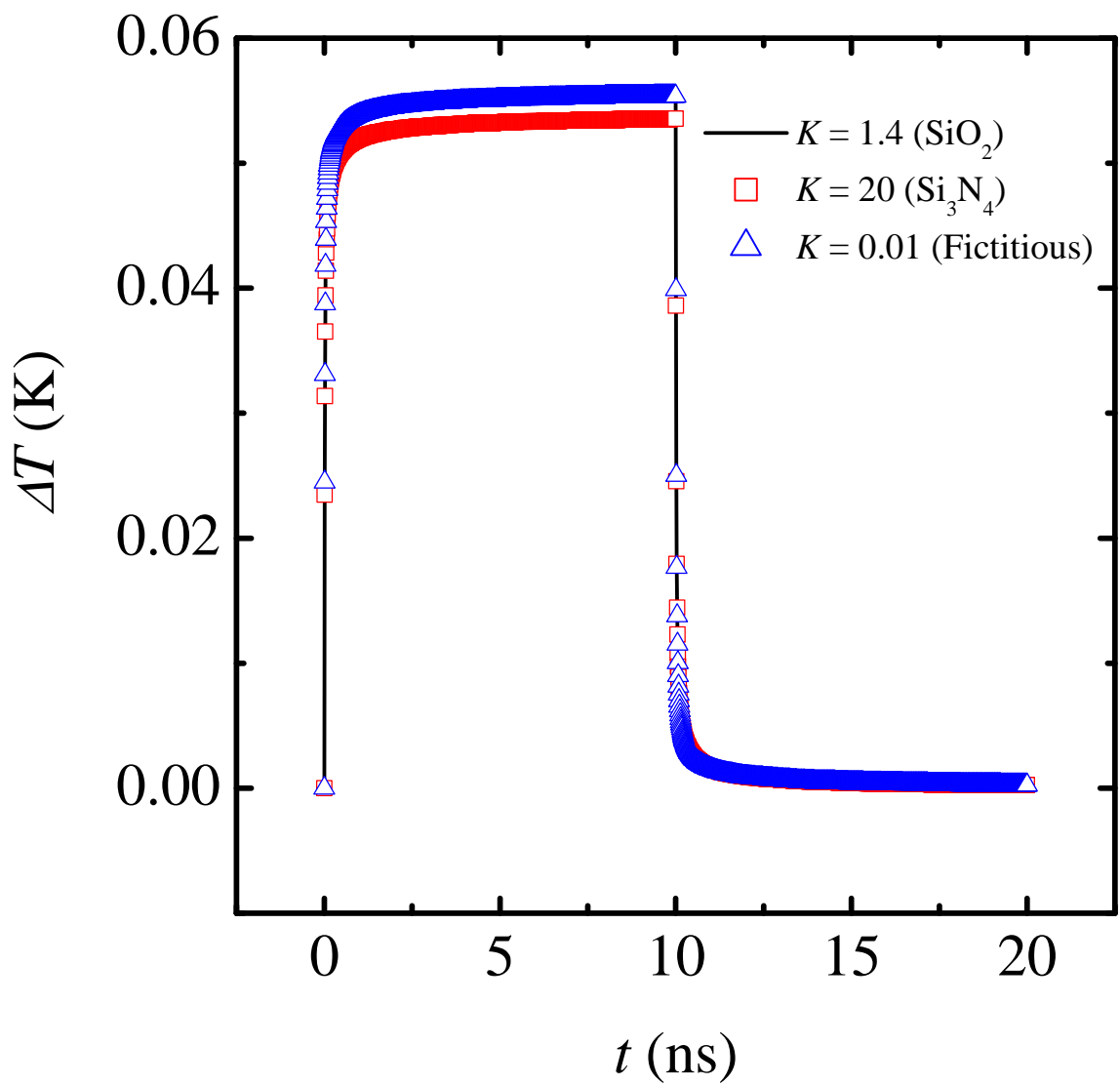

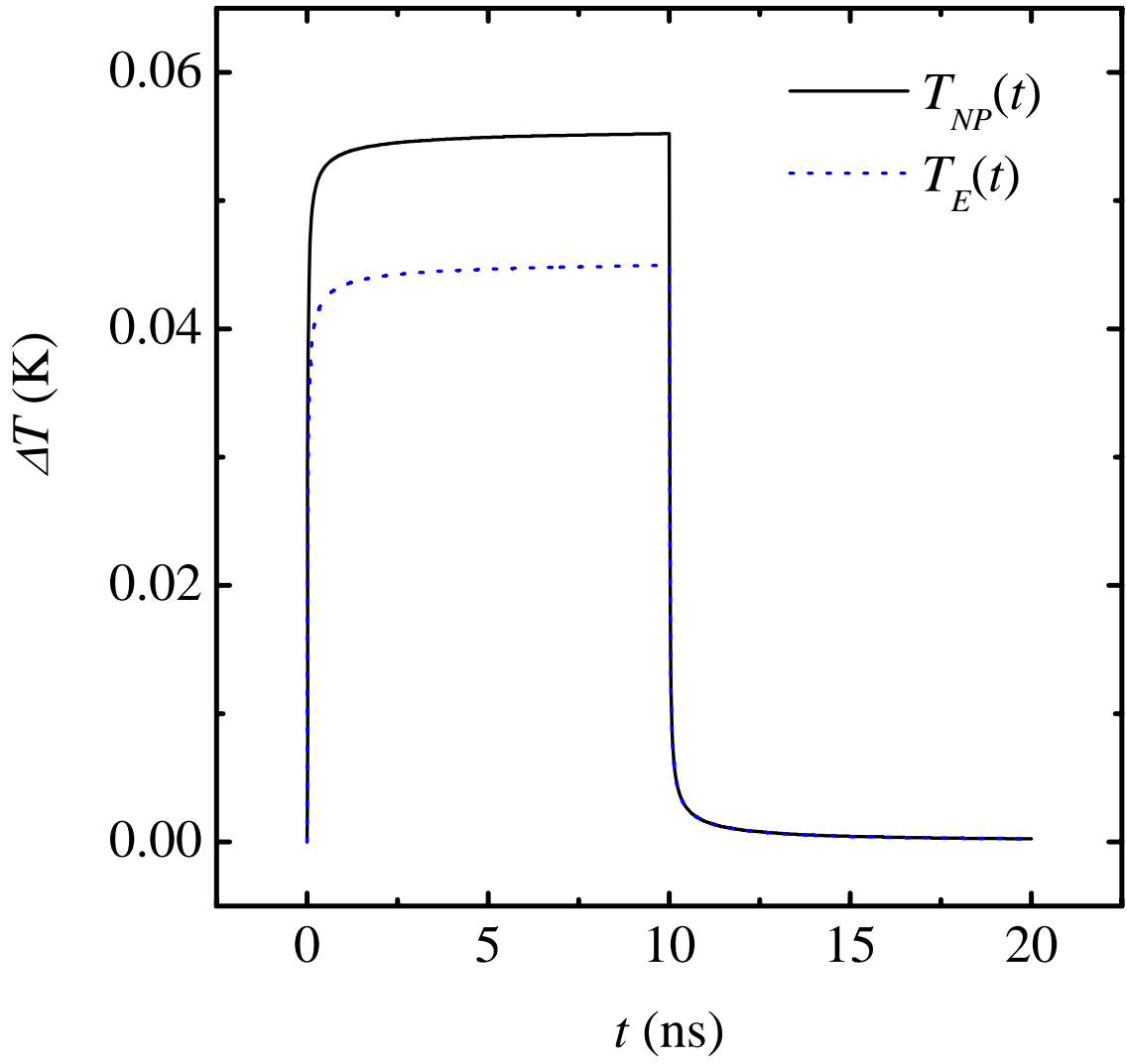

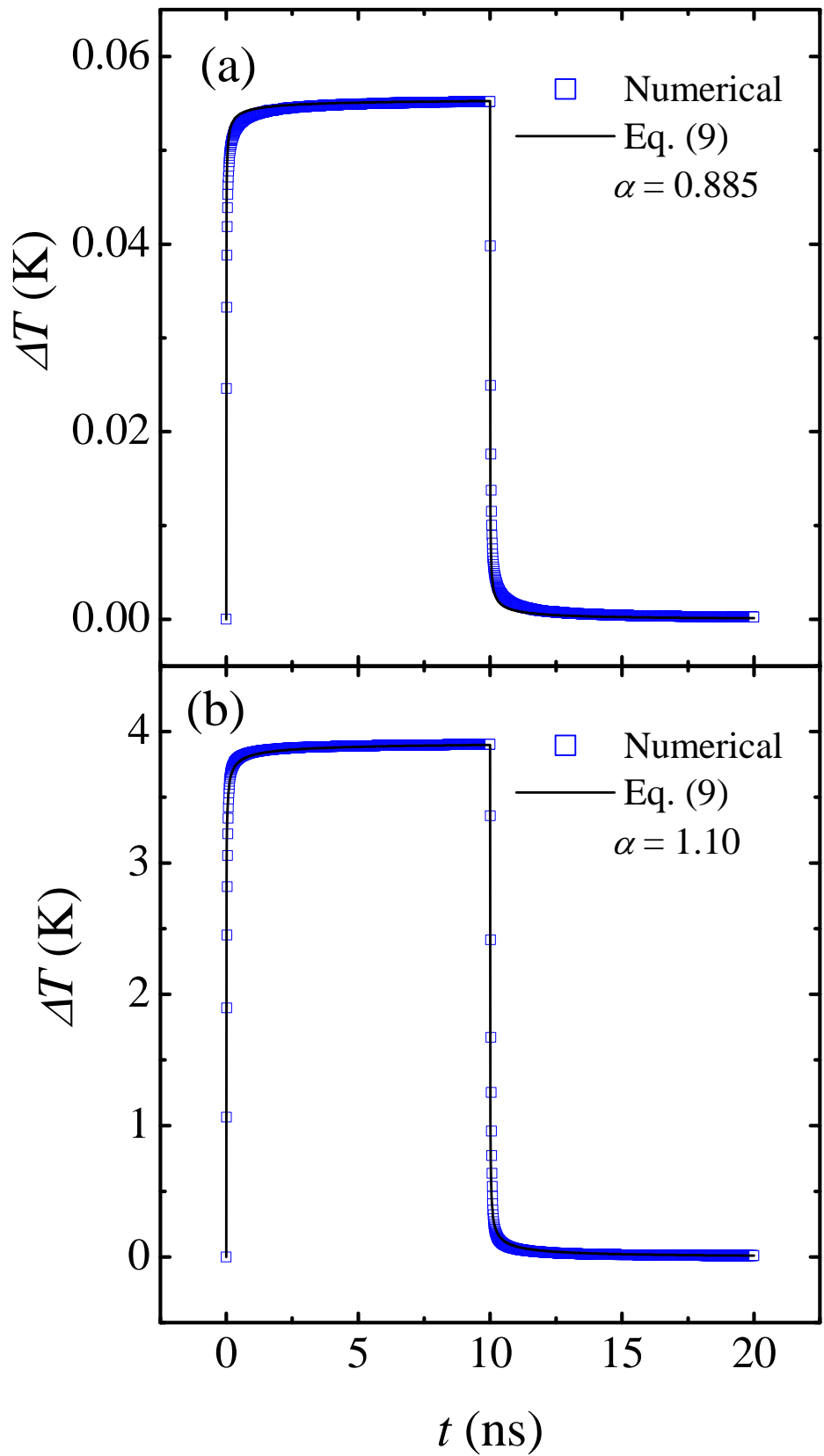

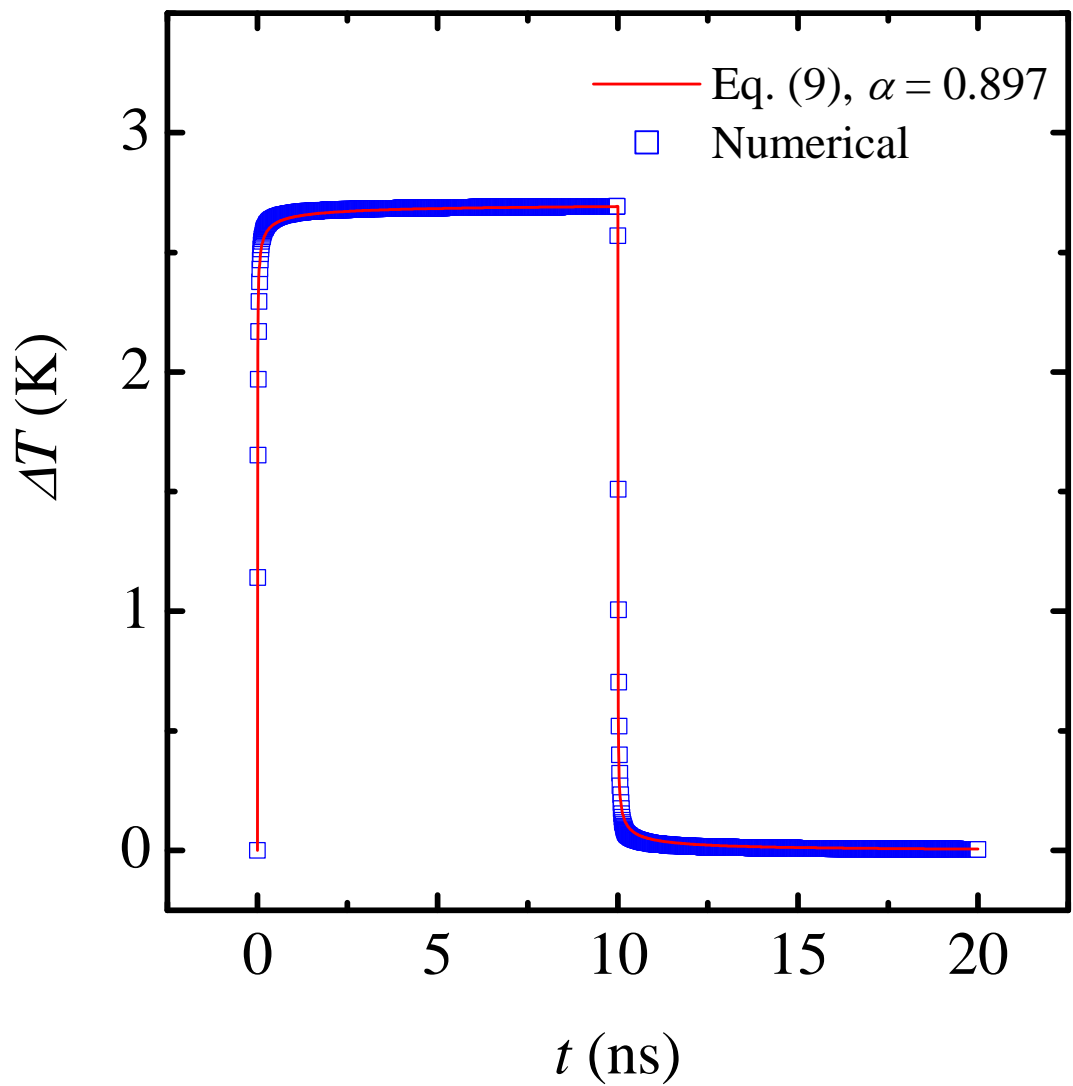